# Improved Spin Dependent Limits from the PICASSO Dark Matter Search Experiment

M. Barnabé-Heider, M. Di Marco, P. Doane, M.-H. Genest, R. Gornea, R. Guénette,
C. Leroy, L. Lessard , J.- P. Martin, U. Wichoski  and V. Zacek

*Département de Physique, Université de Montréal, Montréal, H3C 3J7, Canada*

K. Clark, C. Krauss and A. Noble

*Department of Physics, Queens University, Kingston, K7L 3NG, Canada*

E. Behnke, W. Feighery[*], I. Levine and C. Muthusi[*]

*Departments of Physics & Astronomy, and [*]Chemistry, Indiana University South Bend, South Bend, Indiana, 46634, USA*

S.  Kanalalingam and R. Noulty

*Bubble Technology Industries, Chalk River, K0J 1J0, Canada*

Abstract

The PICASSO experiment reports an improved limit for the existence of cold dark matter WIMPs interacting via spin-dependent interactions with nuclei. The experiment is installed in the Sudbury Neutrino Observatory at a depth of 2070 m.  With superheated $C_4F_{10}$ droplets as the active material, and an exposure of 1.98±0.19 kgd, no evidence for a WIMP signal was found. For a WIMP mass of 29 GeV/$c^2$, limits on the spin-dependent cross section on protons of $\sigma_p$ = 1.31 pb and on neutrons of $\sigma_n$ = 21.5 pb have been obtained at 90% C.L. In both cases, some new parameter space in the region of WIMP masses below 20 GeV/$c^2$ has now been ruled out. The results of these measurements are also presented in terms of limits on the effective WIMP-proton and neutron coupling strengths $a_p$ and $a_n$.

*E-mail address:* zacekv@lps.umontreal.ca (V. Zacek)

## 1. Introduction

PICASSO is an experiment searching for cold dark matter through the direct detection of weakly interacting massive particles (WIMPs) via their spin-dependent interactions with nuclei. It uses a superheated fluorocarbon, $C_4F_{10}$, as the active material and searches for WIMP interactions on $^{19}F$. This choice is motivated by the fact that $^{19}F$ is one of the most favorable nuclei for direct detection of spin-dependent interactions [1, 2, 3]. In fact the nuclear form factor of $^{19}F$ enhances the signal by nearly an order of magnitude compared to other frequently used materials (Na, Si, Al, Cl). In addition, fluorine is complementary to $^{23}Na$ and $^{127}I$, if one expresses the relative sensitivities of the target nuclei in terms of the WIMP-proton ($a_p$) and WIMP-neutron ($a_n$) couplings due to the negative sign in the ratio of the expectation values of the proton and neutron spins [4,5].

To detect dark matter, PICASSO exploits the well established and extremely successful bubble chamber technique [6, 7, 8]. In this technique, the detector medium is a metastable superheated liquid, and as described by Seitz [9], a phase transition is initiated by a heat spike due to the energy $E_{dep}$ deposited by a charged particle traversing a certain critical length $L_c$: $E_{dep} = (dE/dx) \cdot L_c$. At a given temperature a certain critical amount of energy $W_c$ has to be deposited within $L_c$. The quantities $W_c$ and $L_c$ are functions of the surface tension and superheat of the liquid, where superheat is defined as the difference between the vapor pressure of the liquid and the externally applied pressure. At a given ambient pressure both, $W_c$ and $L_c$ decrease strongly with increasing temperature and can be approximated over a large temperature range by $W_c(T) = W_{c0} \exp(-\alpha T)$ and $L_c(T) = L_{c0} \exp(-\beta T)$, where the precise values for the constants have to be inferred e.g. from neutron calibration experiments.

The phase transition criteria $E_{dep} > W_c$ implies that the detection process depends not only on the temperature and the ambient pressure of the superheated liquid, but also on the specific energy loss of the interacting particle. This allows an efficient suppression of backgrounds from particles which produce only low ionization densities, such as cosmic ray muons, γ- and β-rays, while maintaining full efficiency for detection of strongly ionizing nuclear recoils, such as WIMP induced nuclear recoils [10].

By dispersing the superheated liquid in the form of droplets (with diameters between 10 to 100 μm) in polymerized water based gels or viscous liquids, the detectors have essentially a 100% duty cycle, in contrast with conventional bubble chambers. Moreover a phase transition, or droplet burst, is accompanied by an acoustic signal, which is recorded by piezoelectric transducers mounted on the container's exterior wall. Detectors of this type are known as bubble detectors (BD) or superheated droplet detectors (SDD) and have matured into standard devices in neutron dosimetry [11, 12].

Droplet detectors are threshold counters where each individual superheated droplet acts as an independent bubble chamber detector. The threshold depends on temperature and pressure. At higher temperatures, the energy threshold for nuclear recoils is lower. The precise dependence of the recoil energy threshold on temperature and pressure was determined through extensive calibration runs with mono-energetic neutron beams at the Montréal tandem accelerator facility [10,13]. It was found that the efficiency to detect nuclear recoils with energy $E_r$ at a given temperature is not a step function, but rises gradually from threshold to full efficiency:

$$\varepsilon(E_r, E_{th}(T)) = 1 - \exp\left[-a_0 \frac{E_r - E_{th}(T)}{E_{th}(T)}\right] \qquad (1)$$

where $E_{th}(T)$ is the minimum detectable recoil energy determined from neutron calibrations and $a_0 = 1.0 \pm 0.1$ is a free parameter obtained from a fit to the data.

Fig. 1 shows the temperature dependence of the energy threshold $\varepsilon(E_r, E_{th}(T))$ for $^{19}$F recoils in superheated $C_4F_{10}$ for two detection efficiencies, $\varepsilon = 50\%$ and $\varepsilon = 90\%$, and at ambient pressure (1.23 bar at the depth of the underground laboratory). Since PICASSO detectors are operated in the temperature range from 20° to 47°, this translates into a sensitivity to $^{19}$F recoils from 500 to 6 keV ($\varepsilon$ = 50%). WIMP induced recoil energies are expected to be smaller than 100 keV and therefore become detectable above 30°C.

For comparison, Fig. 1 indicates also the regions of sensitivity for other processes in superheated $C_4F_{10}$: *i)* alpha particles from U/Th contaminations in the polymer surrounding the droplets are emitted with energies of typically 5 MeV. Following slow down in the gel, they can reach a droplet with energies from several MeV down to a few keV and depending on the specific energy loss rates of the alpha particles in the droplet, the energy depositions within $L_c(T)$ cover a large range. Therefore alpha counting starts at around 15° C, where only alpha particles with maximum stopping power are able to trigger a phase transition; at higher temperatures the liquid becomes sensitive to alpha particles with increasingly smaller *dE/dx*. *ii)* above 55°C, $^{19}$F recoils with energies below one keV can be detected, but at the same time the detector becomes sensitive to low energy secondary electrons emitted following interactions of x-rays, γ-rays and minimum ionizing particles and the detector loses its discrimination power. *iii)* at the foam limit, thermal fluctuations can trigger phase transitions and the detector becomes intrinsically unstable.

A complete simulation of the detector response to neutrons, alpha particles and gamma-rays was performed and the results were compared to measurements with an Ac(Be) neutron source, n-beams, gamma-ray sources and detectors spiked with alpha emitters ($^{241}$Am, $^{238}$U). The simulations took into account the respective detector-source geometry, the precise geometry and chemical composition of the detectors, the experimentally known distribution of droplet diameters and different energy loss processes [14]. After tracking, the energy depositions by the various particles in the droplets were calculated and transformed into a temperature response using the Seitz theory. The results reported in [15, 16] indicate, that all data can be well described with a unique and consistent set of variables ($W_c$, $L_c$, $\alpha$, $\beta$) which parameterize the underlying model of recoil energy threshold and energy deposition, as predicted by the Seitz model.

Previous results on the application of the superheated droplet technique in dark matter searches have been published by the PICASSO and SIMPLE groups [10, 17].

2. Detectors and Experimental Set Up

The PICASSO experiment is installed at a depth of 2070 m in the Sudbury Neutrino Observatory. At this location the cosmic muon flux is reduced to less than 0.27 muons/m$^2$/day [18]. The ambient thermal neutron flux from the rock was measured to be 4144.9 ± 49.8 ± 105.3 neutrons/m$^2$/day [19]. The fast neutron flux is less well known, but has been estimated to be about 4000 neutrons/m$^2$/day [18]. The set up used in this analysis consists of 3 detector modules of 1.5 liter volume each in cylindrical containers of 9 cm diameter and 35 cm height. The containers were fabricated from Himont SV258 polypropylene (selected for its low radioactivity), and had stainless steel lids sealed with a polyurethane O-rings. The $^{222}$Rn emanation rate for these containers was measured to be only 15 ± 9 atoms/day [20]. The detectors were fabricated in a clean environment and special efforts were made to purify the detector components to reduce alpha particle emitters from the U and Th decay chains. The ingredient that contributes most is CsCl, a heavy salt which was dissolved at a concentration of 1.62 g/ml into the water of the gel matrix in order to match the density of the gel solution to that of the droplets before polymerization. The CsCl and all other gel components were purified using HTiO ion exchange techniques developed by the SNO collaboration [21] down to an assayed level of 2.8 ± 0.7 x 10$^{-10}$gU/g [20]; the active detector liquid was distilled once.

Each detector is filled with 1 litre of polymerized emulsion loaded with droplets of $C_4F_{10}$ with an average diameter of 11 μm. The loading, i.e. the fraction of active material in the detectors was determined in several ways: a) by scanning many samples of a detector under a microscope; b) by precisely weighing all the detector components before and after detector fabrication; c) by calibrating test detectors in a known flux of a mono-energetic neutron beam; d) by placing detectors with known active mass determined from a), b) and c) at a fixed distance in the flux of an Ac(Be) source emitting 1.31 x $10^5$ ± 3 n/s, their response in cts/g·n·cm$^{-2}$ was used as a reference for the calibration of detectors with unknown loading; e) by comparing the measured rates in d) with Monte Carlo simulations, using the measured droplet size distribution and neutron flux as input and leaving the loading as a free parameter. Results from all methods were in agreement. A typical detector loading is around 0.5% (i.e. g of $C_4F_{10}$ / g of gel). For the detectors used in this analysis, the loading was determined by a fast neutron calibration (method d) with an error of 8%, dominated by statistics.

Each detector is read out by two piezoelectric transducers (Phys. Acoustics, DTI5MT): one sensor is glued at the center of the outer wall of the container; the other is mounted on the opposite side, with a 5 cm offset in height. The transducers are adapted to the sound emission spectrum with a frequency response from 20 kHz to 1 MHz. The signal amplitude depends on droplet size, temperature, solid angle and sound attenuation, and its duration is typically several ms long. The signal is amplified by a gain of 5000 and fully digitized in a 1 MHz FADC. An event triggers if either transducer signal is larger than a threshold of 500 mV, in which case both channels are read out. The trigger acceptance is a function of temperature and has been obtained from an extrapolation of the signal amplitude distribution below threshold. It has been determined to be $\varepsilon_t$ = 0.95 ± 0.05 in the range of 25 °C to 45 °C, decreasing to $\varepsilon_t$ = 0.88 ± 0.10 at 20 °C. The dead time of the acquisition system is negligible.

A batch of three detectors was installed in a thermally and acoustically insulated box. The temperature was controlled by two Peltier junctions and four temperature sensors, one glued on each of the detector walls, and one located inside the box to monitor the air temperature. The absolute temperature in the unit was known with a precision of ±1.5 °C and was stabilized within ±0.1 °C.

The uncertainty quoted for the temperature is a consequence of temperature gradients within the detector volume combined with the uncertainty in the temperature calibration. A layer of acoustically insulating foam was installed below each module. Two temperature-controlled units were run in parallel and water cubes, providing a 33 cm thick shielding against neutrons coming from the rock walls, surrounded tightly the entire set up.

Data taking was organized into series of runs at a particular temperature. Runs were normally of 30 h duration although in three of a total of thirteen series there were runs of longer duration (60h and 2 x 90h). After each run, the detectors were compressed with nitrogen and kept at a pressure of 8 bar for 10 hours in order to recompress bubbles to droplets and to prevent bubble growth, which could damage the polymer surrounding the droplets. In this mode of operation, an overall duty cycle of 80% can be obtained. The detector and the compression system were required to be radon tight, so the system was hermetically sealed and checked using a helium leak detector.

The long term stability has been measured for several detectors and the rates were found to be stable to better than 11% by measuring the sensitivity to fast neutrons repeatedly and over intervals of more than one year. This is an upper limit and obtained from detectors heavily exposed to radiation during neutron calibration tests. The observed loss in sensitivity has been attributed to a fracturing of the gel and loss of active mass occurring close to the top surface of the detector due to $N_2$ infiltration into the gel matrix during recompression. In the future using a hydraulic recompression system will eliminate this effect.

3. Measurements

The set up at SNO was installed in July 2002 and served as a test facility to study all performance aspects in view of a larger PICASSO experiment. Many detectors with decreasing level of intrinsic contamination and background have been studied since. The data set used in this analysis includes only those detectors that have passed through the full purification process. These data were recorded between April 2004 and October 2004 using detectors SBD40, SBD46 and SBD47. The detectors had an active mass of $^{19}$F of 7.45 ± 0.67g, 6.62± 0.60g and 5.35 ± 0.48g, respectively. All detectors were calibrated in the same known neutron field of an AcBe source before being installed underground. The temperatures were varied during these runs in a random sequence from 20 to 47 °C, covering an energy range of $^{19}$F recoils from 6 keV to 500 keV. During the entire measurement period data taking proceeded continuously under remote control via Internet from Montréal.

The data analysis treated each detector independently. Each signal waveform was transformed into the frequency domain using a fast Fourier transform (FFT). All events had to pass an initial filter which rejected non-droplet like background signals by performing a wave-form analysis on the FFT. This filter exploited the fact that the frequency for most external noise sources of acoustic or electromagnetic origin was below 20 kHz, while particle induced events have a substantial component in a band from 20 to 130 kHz. The selection efficiency of the algorithm has been studied and evaluated with high statistics in runs with count rates of about 1 Hz with fast neutrons from an AcBe source and with dedicated calibration detectors spiked with alpha- particle emitters ($^{241}$Am, $^{238}$U). The acceptance was found to be uniform over the temperature range from 25 °C to 45 °C with $\varepsilon_f$ = 0.85 ± 0.05, decreasing slightly for lower temperatures.

Time cuts were then applied to ensure that the analysis included only data taken when the detectors were in a stable operating configuration. In order to guarantee a stable and uniform temperature distribution inside the detectors, we excluded the first 30h long run, recorded after a temperature change. The detectors were often observed to produce a few sporadic events shortly after being decompressed. The origin of this effect was understood after tests were performed with detectors without active gas. These events were due to the formation of gas bubbles in the gel matrix, a consequence of $N_2$ gas having diffused into the gel during the recompression cycle. To eliminate these noise events from the analysis, 2 hours were cut from the data stream following each recompression period. Finally, as a precaution, multiple events were removed if they occurred within 10 minutes of a primary event, since these secondary events might have been triggered by a weakening of the gel matrix in the proximity of the primary event. The primary event was retained. This effect is small (cutting less than 0.5% of the events), is absent for SBD40, and increases the dead time by only 2.6%.

4. Results and Discussion

The effective exposure for the ensemble of detectors SBD 40, 46 and 47 amounts to 1.98 ± 0.19 kg-days of $^{19}$F after time cuts, where the error is given by the 8% uncertainty on the determination of the active mass of the detectors and a 5% statistical error from the determination of the filter efficiency. The count rates for the three detectors, corrected for efficiencies and acceptances, and plotted as a function of temperature, are shown in Fig. 2. No correlations were detected between events of different detectors, nor were correlations found with seismic or mine related activities [22]. For the cleanest detector, SBD40, the highest count rate, obtained at the highest temperature, is about 480 counts per kg of $^{19}$F per day. This represents the integrated background over the entire recoil spectrum above a threshold of 6 keV.

The observed data are well described by the temperature profile of the alpha response of the detectors. The shape of the alpha response has been determined with high precision in independent measurements with $^{238}$U- and $^{241}$Am- doped detectors and is shown in the insert of Fig. 2. After correction for the filter acceptance, the experimental response is equally well reproduced by Monte Carlo simulations using the same parameters, which describe the detector response to mono-energetic test beam neutrons [16]. The fit of the alpha curve alone to the data results in a reduced chi-squared of 1.15, 1.7 and 1.4 for the detectors SBD40, 46 and 47, respectively, for 10 degrees of freedom.

Several effects have been studied which could possibly influence the shape of the alpha response curve. No significant change was found in simulations when varying the alpha particle energies within the range of energies occurring within the U and Th decay chains. Since the droplet size distributions can differ slightly between different detectors, simulations were performed by varying the mean of the droplet size distribution within the range, which might occur during detector fabrication. Shape effects near threshold are expected, since by geometric considerations larger droplets have a higher geometric probability to sample the large *dE/dx* values at the end of the alpha track, than do smaller droplets [23]. Thus a larger average droplet size lowers the alpha detection threshold temperature and the shape of the response curve becomes sensitive to the droplet size distribution in its vicinity. The simulations have shown that these shape variations are smaller than 5% in the interesting range between 25°C and 45°C.

In order to construct the response curve for WIMP- induced recoils we follow the recommendations of [24] and use as halo parameters a local WIMP matter density of 0.3 GeV/c$^2$/cm$^3$, a neutralino velocity dispersion in the halo of 230 km/s, an earth- halo relative velocity of 244 km/s and an escape velocity of 600 km/s. Keeping the WIMP mass and cross section as parameters, the expected WIMP induced count rate is then calculated as a function of temperature by applying the corresponding detector threshold function (Eq. 1) to the expected WIMP induced recoil spectrum. The expected WIMP response curves differ significantly from that of the alpha background, the latter extending towards much lower temperatures. Therefore the two contributions can be fit simultaneously to the data and in case of no significant deviation from the alpha-background, an upper bound can be obtained for WIMP induced recoils as a function of the WIMP interaction cross section on $^{19}$F ($\sigma_F$) and the WIMP- mass (M$_{WIMP}$). Curve fitting was done with MINUIT by using a two parameter fit to minimize the global chi-squared. The two free parameters in the fit were the scale factor of the known alpha response curve and the cross section for a given WIMP mass.

The general spin dependent cross section for WIMP scattering on $^{19}$F has the form [3,4,5,24]:

$$\sigma_F = 4G_F^2 \mu_F^2 C_F \qquad (2)$$

where $G_F$ is the Fermi coupling constant and $\mu_F$ is the WIMP- $^{19}$F reduced mass. $C_F = (8/\pi)\left(a_p \langle S_p \rangle + a_n \langle S_n \rangle\right)^2 (J+1)/J$ is a spin-dependent enhancement factor, where $a_p$, $a_n$ are the effective proton and neutron coupling strengths, $J = 1/2$ is the total nuclear spin, and $\langle S_p \rangle = 0.441$ and $\langle S_n \rangle = -0.109$ are the expectation values of the proton and neutron spins within $^{19}$F [25].

In order to compare our results with other experiments [24], we convert the cross section limits on $^{19}$F into limits on the WIMP-proton ($\sigma_p$) and WIMP neutron ($\sigma_n$) cross sections, respectively. To do this, we follow the procedure outlined in [3] by assuming that all events are either due to WIMP-proton or WIMP-neutron elastic scatterings in the nucleus, i.e $a_n = 0$ or $a_p = 0$, respectively, and obtain:

$$\sigma_i = \sigma_F \left(\frac{\mu_i^2}{\mu_F^2}\right) \frac{C_i}{C_{i(F)}} \qquad (3)$$

where $i$ = proton or neutron, $\mu_i$ is the WIMP-nucleon reduced mass (the mass difference between neutron and proton is neglected) and $C_{p(F)}$ and $C_{n(F)}$ are the proton and neutron contributions to the total enhancement factor of $^{19}$F. They are related to the $a_p$ and $a_n$ couplings by $C_{i(F)} = (8/\pi) a_i^2 \langle S_i \rangle^2 (J+1)/J$. $C_p$ and $C_n$ are the enhancement factors for scattering on individual protons and neutrons. The values for the ratios $C_{p(F)}/C_p = 0.778$ and $C_{n(F)}/C_n = 0.0475$ are taken from [25].

Fig.3 compares various neutralino-proton cross sections for different neutralino masses with the measured temperature profile of our cleanest detector, SBD40, and for comparison, a detector of an earlier generation, SBD32, which was not produced in a clean environment. The fit of the alpha curve to the data of SBD32 yields a reduced chi-squared of 0.7.

By combining the fit results of the three detectors SBD40, 46 and 47 in a weighted average, we obtain maximum sensitivity for a WIMP mass of $M_{WIMP}$ = 29 GeV/c$^2$ and a cross section of $\sigma_p$ = 0.20 ± 0.80 pb (1$\sigma$), which is compatible with no effect. We translate this result into an upper bound on the spin dependent cross section on protons of $\sigma_p$ = 1.31 pb at 90% C.L and $M_{WIMP}$ = 29 GeV/c$^2$. Alternatively, if the ensemble of the three detectors is treated as one detector and the three data sets are simply added together, we obtain a similar result, with a limit of $\sigma_p$ = 1.36 pb at 90% C.L and $M_{WIMP}$ = 29 GeV/c$^2$. The resulting exclusion plot of cross section as a function of WIMP mass, shown in Fig. 4a, is adopted from ref. [5] and references therein and compares our result with the most recent results in the spin-dependent sector of WIMP interactions on protons.

Although the nuclear spin in $^{19}$F is mostly carried by the unpaired proton, the non-vanishing neutron enhancement factor $C_n$ also gives our experiment some sensitivity in the neutron sector. The sensitivity to scattering on neutrons is reduced by the factor $(C_n/C_{n(F)})/(C_p/C_{p(F)})$ = 0.061 with respect to protons and translates into a limit on the WIMP- neutron cross section of 21.5 pb at 90% C.L. and $M_{WIMP}$ = 29 GeV/c$^2$. The resulting exclusion plot is shown in Fig. 4b together with a summary of existing limits from other experiments. In both cases, in the proton and the neutron sector, some parameter space in the region of small WIMP masses below 20 GeV/c$^2$ is ruled out by PICASSO, which is not yet excluded by other experiments.

From the WIMP-proton and WIMP-neutron elastic scattering cross section limits (Fig. 4) one finds the allowed region in the $a_p$-$a_n$ plane from the condition [3]:

$$\left(\frac{a_p}{\sqrt{\sigma_p}} \pm \frac{a_n}{\sqrt{\sigma_n}}\right)^2 < \frac{\pi}{24 G_F^2 \mu_p^2} \qquad (4)$$

where the relative sign inside the square is determined by the sign of $\langle S_p \rangle / \langle S_n \rangle$. In our experiment, the values of $a_p$ and $a_n$ are constrained, in the $a_p$-$a_n$ plane, to the inside of a band defined by two parallel lines of slope $-\langle S_n \rangle / \langle S_p \rangle = 0.247$.

Fig. 5 shows the excluded region, which lies outside the band in the $a_p$-$a_n$ plane for the PICASSO experiment compared to other experiments for $M_{WIMP} = 50$ GeV/c$^2$.

The following uncertainties (1σ) have been included in the evaluation of these limits: *a)* a 9.4 % error on the effective exposure (active mass of $^{19}$F × measuring time), mainly given by an 8% error on the determination of the active mass of the detectors and a 5% statistical error from the determination of the filter efficiency; *b)* a 5% uncertainty in long term detector stability; *c)* a 5% uncertainty on the temperature dependence of the trigger acceptance; *d)* a 7% error in the neutralino response curve, mainly due to an uncertainty of ± 2 °C in the knowledge of $T_c$, the critical temperature of $C_4F_{10}$ which enters in the energy threshold applied to the spectrum of neutralino induced recoils; *e)* a 5% error due to possible distortions of the alpha response curve towards lower temperatures, introduced by uncertainties in the droplet size distribution; and *f)* an error of ±1.5 °C in the temperature calibration.

5. Conclusions

The PICASSO experiment has established a set of low background detectors at the Sudbury Neutrino Observatory to search for spin-dependent WIMP induced interactions on $^{19}$F using the superheated droplet technique. The results from three superheated droplet detectors with a total active mass of 19.4 ± 1.0 g of $^{19}$F and an exposure of 1.98 ± 0.19 kgd were analysed. No positive evidence for WIMP induced nuclear recoils has been obtained and a 90% C.L. upper limit of 1.31 pb on protons and 21.5 pb on neutrons was obtained for a WIMP mass $M_{WIMP} = 29$ GeV/c$^2$. This result is based on the absence of events above the known detector background. The sensitivity of the experiment is presently limited by alpha emitting contaminants and its small active mass. The next step of the experiment is in preparation, with 1 kg of active mass, detector modules of 4.5 litre volume, hydraulic recompression, event localization capability and improved purification techniques.

Acknowledgements

The support and hospitality of the Sudbury Neutrino Observatory and its staff are gratefully acknowledged. The skilful technical support of J. Berichon and G. Richard (UdeM) during all phases of the project was very much appreciated. Many thanks go to G. Azuelos (UdeM) for helpful discussions. We thank J.-L.Vuilleumier and J. Busto for performing Ge -assays of detector samples in the Neuchâtel low background laboratory and J. Bocan, S. Pospisil and I. Stekl (Prague) for measuring the Rn diffusion in PICASSO container walls. We thank T. Girard and F. Giuliani (Lisbon) for the drawing program used to generate Figure 4. This work has been funded by the Canadian National Science and Research Council (NSERC) and partial support has been obtained from the Laboratory of Advanced Detector Development (LADD) in the framework of the Canada Foundation for Innovation (CFI). The authors from Indiana University South Bend wish to thank the IUSB R&D committee, and the office of the Dean of the College of Liberal Arts and Sciences for partially funding this work.

**Figures:**

Figure 1: Evolution of the energy threshold for $^{19}$F recoils in $C_4F_{10}$ as a function of temperature (1.23 bar). At threshold the $^{19}$F recoil detection efficiency rises gradually and the broken (continuous) diagonal curves indicate detection efficiencies of $\varepsilon$ = 50% and 90%, respectively. The expected maximum WIMP induced $^{19}$F recoil energies are smaller than 100 keV and become detectable above 30°C; at around 15° C the detector becomes sensitive to alpha particles from U/Th contaminations. Above 55°C recoils with energies below one keV can be detected, but at the same time the detector becomes sensitive to γ-rays, minimum ionizing particles due to associated δ- and Auger electrons; at the foam limit the detector becomes intrinsically unstable.

Figure 2: Count rate as function of temperature for three detectors installed underground in SNOLAB. The curve fit to the data is the calibrated alpha response of the detectors loaded with $C_4F_{10}$. The insert shows the alpha response curve obtained in independent measurements with high statistics from calibrations with $^{238}$U and $^{241}$Am spiked detectors.

Figure 3: The expected alpha and neutralino response curves are compared with the data of detector SBD40, which is the cleanest detector under study and detector SBD32, from an earlier stage of the project where fabrication was not yet done in a clean environment. The expected response for WIMP induced nuclear recoils is shown for a WIMP mass of 50 GeV/c$^2$ and cross sections of 2 pb (dash dotted), 5pb (dotted) and 10 pb (broken), respectively.

Figure 4: Recent limits on the spin-dependent WIMP-proton cross section (a) and WIMP- neutron cross section (b) at 90% C.L. as a function of the WIMP mass. The curves labeled PICASSO 2004 and PICASSO refer to the analysis in this paper and an earlier result published in [10]. These plots have been adopted

from ref. [5], with additional experimental data from [26] and references therein.

Figure 5: Excluded regions in the plane of the effective coupling strengths $a_p$, $a_n$ for protons and neutrons [4,5]. The present experiment, labeled PICASSO 2004 is compared to other experiments for a WIMP mass of 50 GeV/c$^2$.

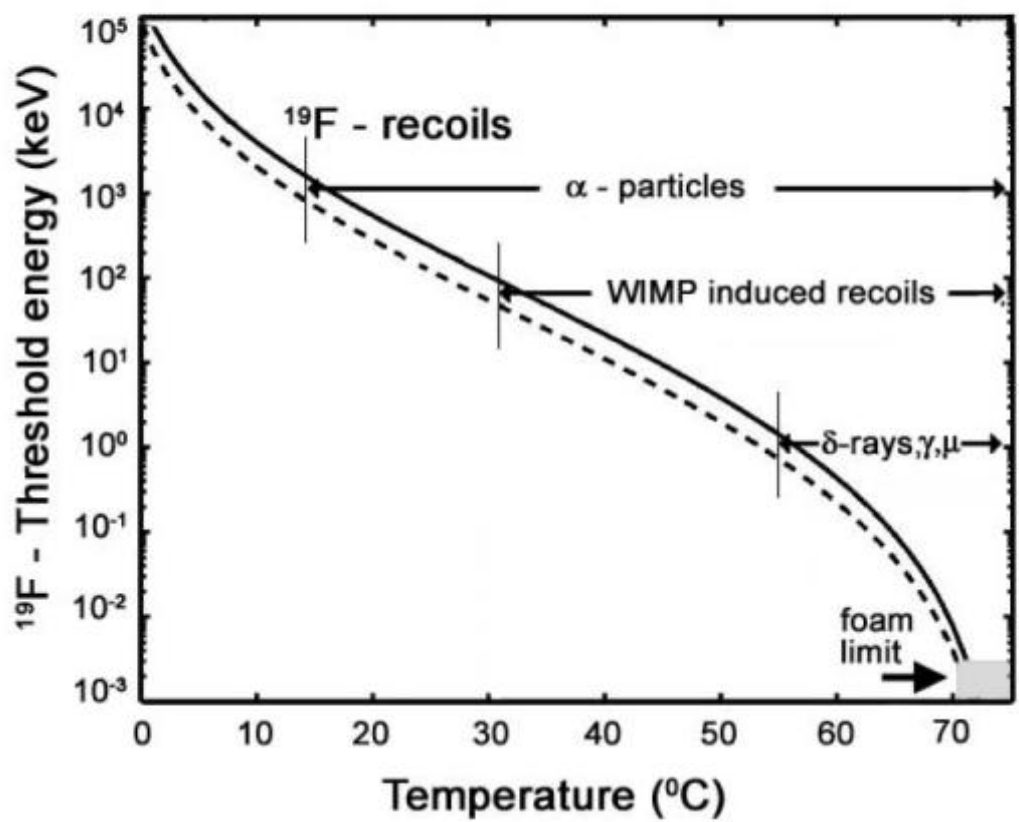

Figure 1

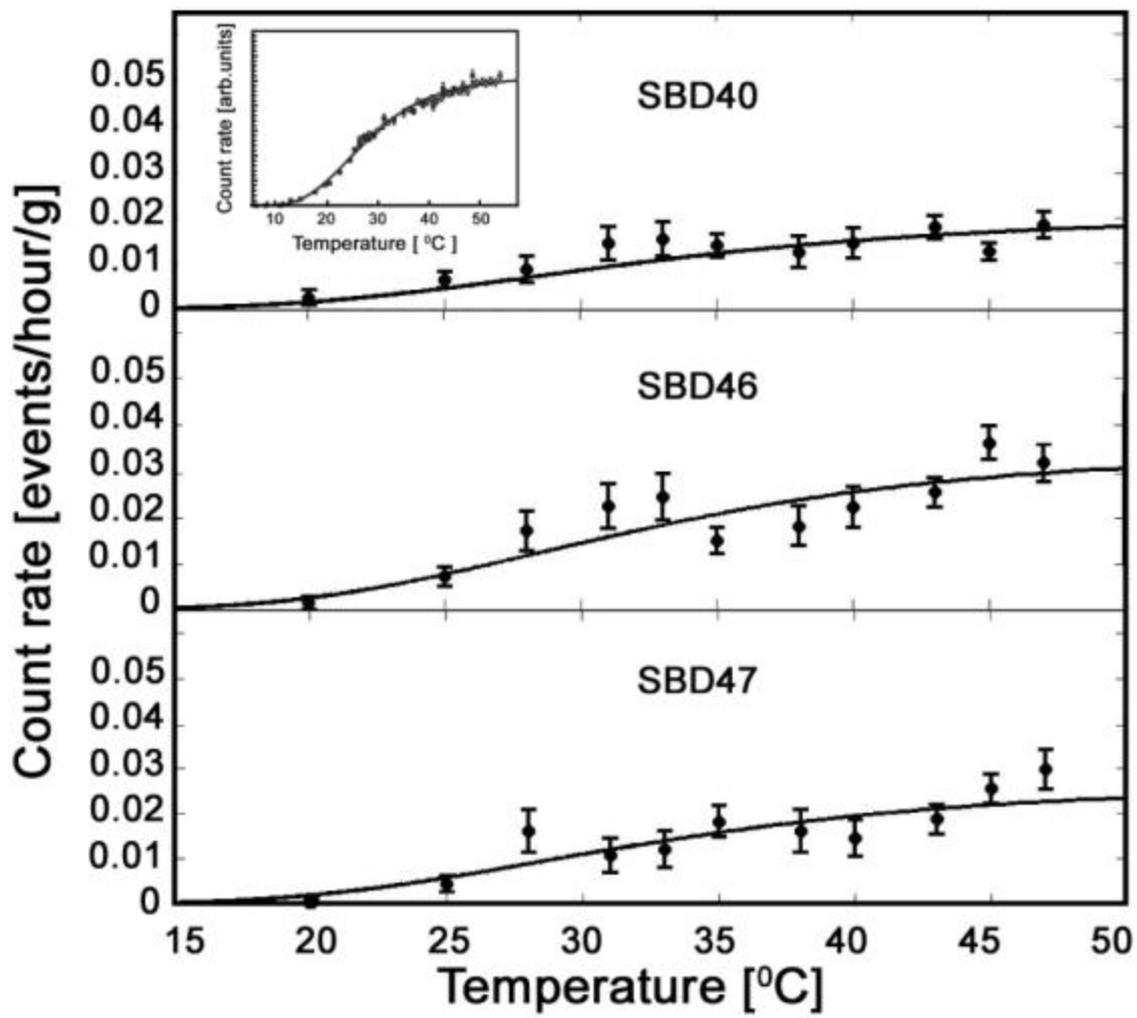

Figure 2

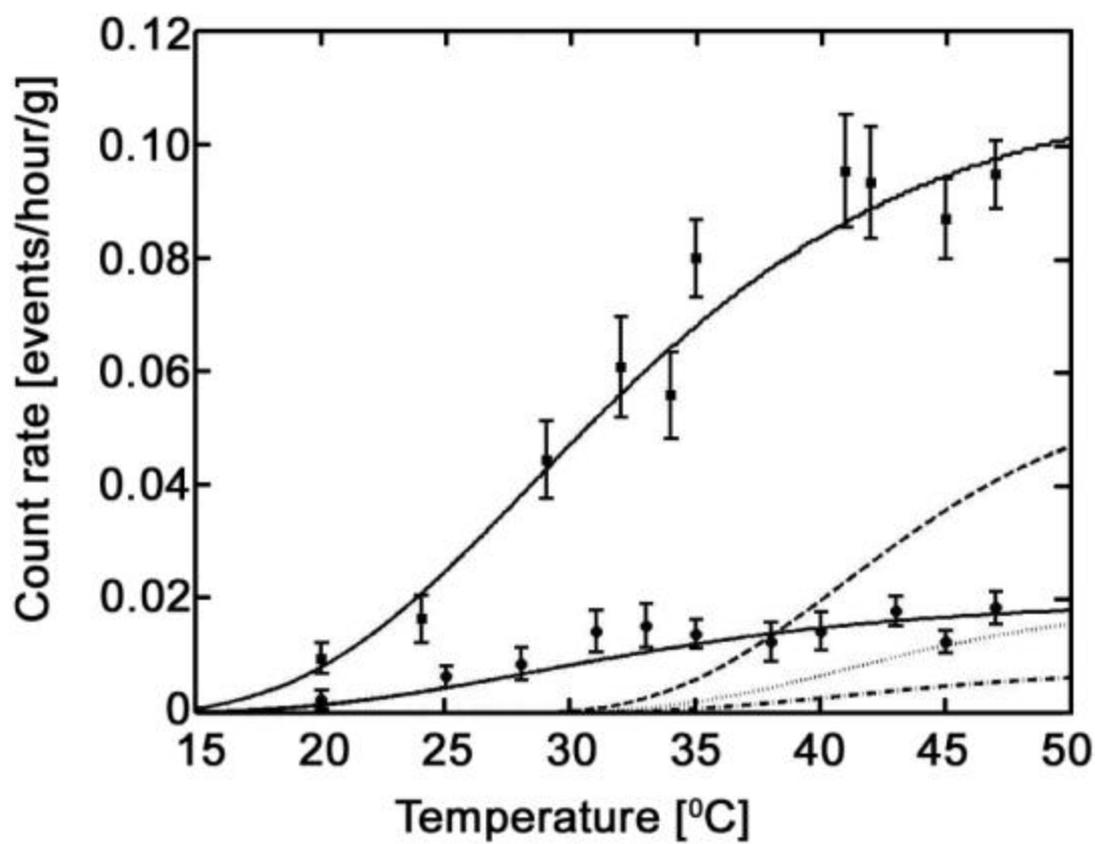

Figure 3

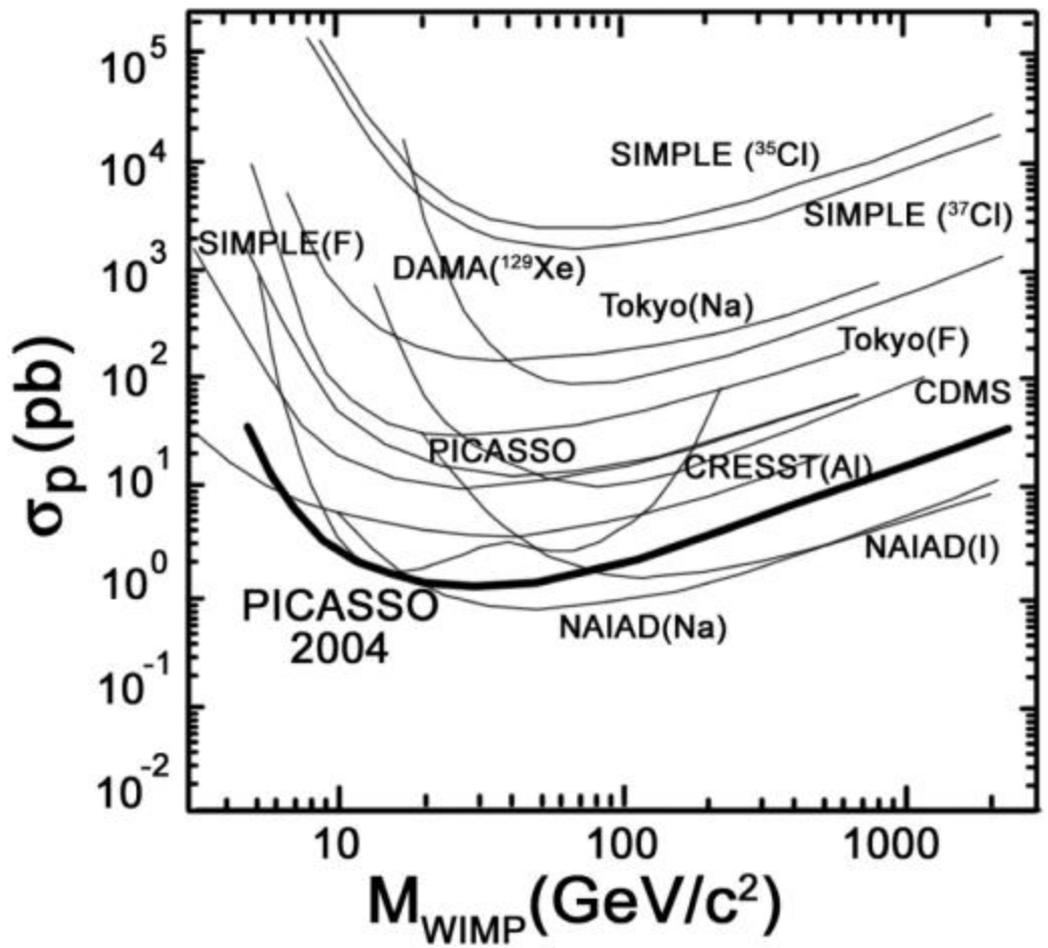

Figure 4a

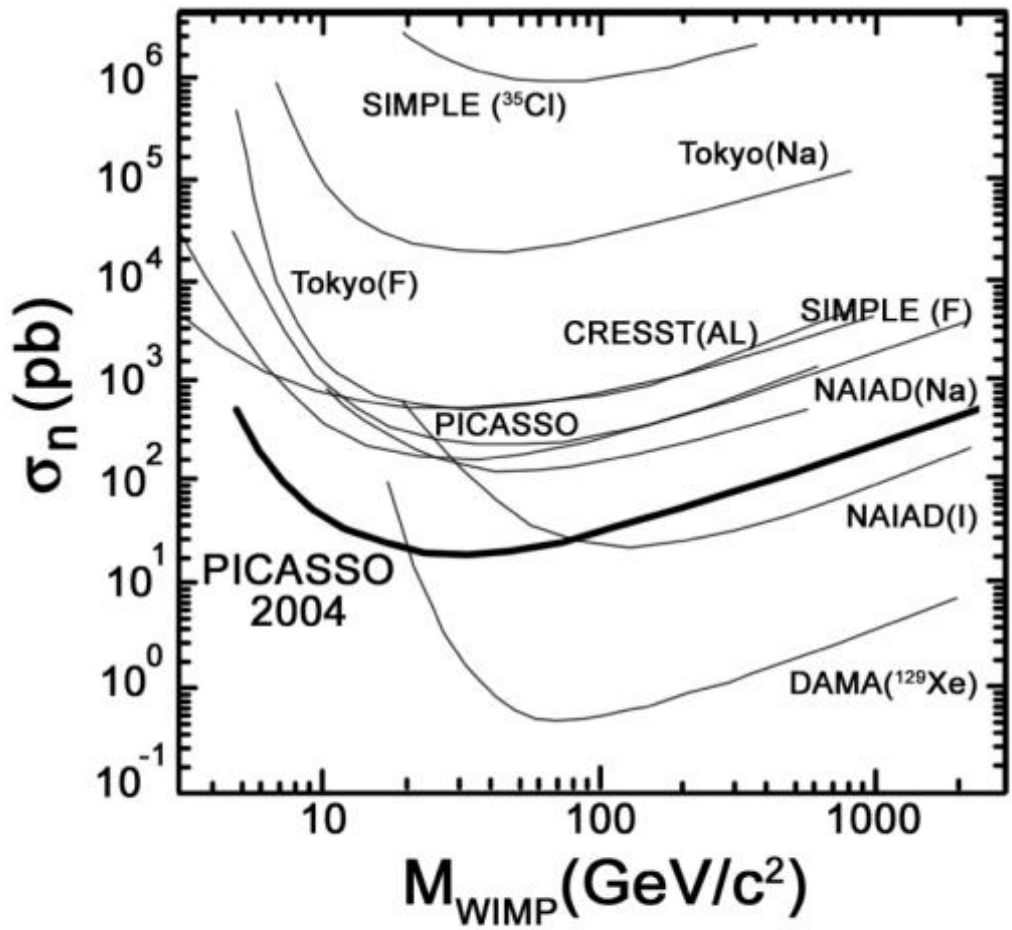

Figure 4b

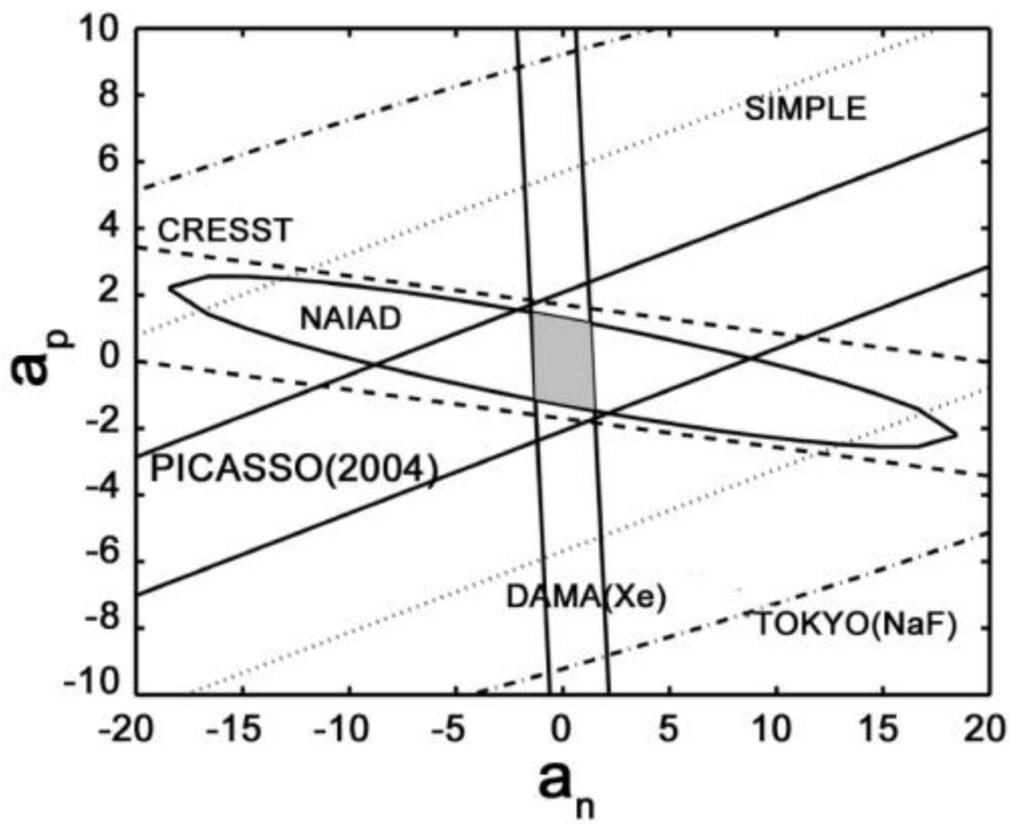

Figure 5